# COMPUTATION OF MEAN VELOCITY DISTRIBUTION IN A TURBULENT FLOW

## A. E. Karpelson

### I. INTRODUCTION

At the time being numerous books and papers are dedicated to theoretical and experimental investigation of a turbulent motion due to its great practical and scientific importance.

By using different experimental methods, we can obtain the velocity profiles in turbulent flows. However, we are not currently able to accurately calculate these distributions. Presently, only special and complex numerical methods can, to some extent, predict aspects of the turbulent flow fields. An analytical determination, at least for some cases, is desirable, because it could lead to a better understanding of the turbulent flow problems. It also provides a correct description of the flows at various boundary conditions and aids in the prediction of flow features.

The Navier-Stokes equations, valid for any turbulent flow, can give us, in principle, the instantaneous velocity and pressure distributions. However we are not usually interested in the "fine structure" of a random chaotic turbulent motion. These instantaneous quantities are always unsteady and depend strongly on the smallest alterations of the initial and boundary conditions, which are never known precisely. Moreover, these quantities are of no practical use at all because of the instability with respect to small disturbances, which always occur in any flow.

What we need is the mean velocity profiles because only such distributions can give a reliable information about main statistical characteristics of a chaotic turbulent flow and can be compared with the experimental data. The mean velocity distributions can be obtained from the modified Navier-Stokes equations, which were devised by Reynolds in 1895. These Reynolds equations were successfully used by many authors for turbulent flow description in general and for velocity field calculation in particular: Boussinesq (the theory of eddy viscosity), Prandtl (the mixing length theory), Taylor (the vortisity transport theory), Von Karman (the similarity hypothesis) and others [1-5].

As a result, there are a number of semi-empirical or empirical solutions for velocity profiles in turbulent flows: the linear law for viscous sub-layer, power law and logarithmic law for the whole flow and their complex modifications [1-5]. All the theories named above were based on various approximations and conjectures aimed at the estimation of one term in the Reynolds equations containing the fluctuating velocity components. As a rule, such theories are valid only for the boundary layer and so their results deviate significantly from experimental data in the central region of a flow, and the solutions do not have a zero derivative at the center of flow, that leads to an unreal broken velocity profile. Moreover, these solutions have a few empirical coefficients, at the time being there is no consensus on the exact value of these constants, which depend on the Re number, roughness of channel wall and other factors.

In general, to express correctly the fluctuating velocity components through the mean velocity, we should have a clear physical understanding of the turbulent process.

Only modern theory of turbulence can give the reasonable explanation of this physical phenomenon.

In this paper rather simple equations and their solutions valid for different turbulent flows are obtained. They describe the smooth mean velocity distribution through the entire flow, satisfy all boundary conditions and specified experimental flow parameters.

## II. GENERAL EQUATIONS

The velocity and pressure distributions in the flow of viscous fluid or gas are described by the Navier-Stokes equations [1-6]

$$\frac{\partial V_i}{\partial t} + V_k \frac{\partial V_i}{\partial x_k} = -\frac{1}{\rho}\frac{\partial P}{\partial x_i} + \nu \frac{\partial^2 V_i}{\partial x_k^2} \,, \tag{1}$$

where $V_i$ are the velocity components, $P$ is the pressure, $x_k$ are the coordinates, $t$ is time, $\nu$ is the kinematic viscosity of fluid or gas, and $\rho$ is its density.

In equation (1) and further we will employ the standard summation convention in which the repeated indices are summed.

In a turbulent flow the total velocity $V_i$ can be represented as a sum of a deterministic function (the mean velocity $\langle V_i \rangle$) and a random function (the fluctuating velocity $\delta V_i$):

$$V_i = \langle V_i \rangle + \delta V_i \tag{2}$$

Substituting (2) into (1) and using ensemble averaging, we obtain the well-known Reynolds equation [1,2,4]

$$\frac{\partial \langle V_i \rangle}{\partial t} + \langle V_k \rangle \frac{\partial \langle V_i \rangle}{\partial x_k} + \overline{\delta V_k \frac{\partial (\delta V_i)}{\partial x_k}} = -\frac{1}{\rho}\frac{\partial \langle P \rangle}{\partial x_i} + \nu \frac{\partial^2 \langle V_i \rangle}{\partial x_k^2} \,, \tag{3}$$

where the third term in left-hand side is the average of product of random functions of two fluctuating velocity components.

Using the equation of continuity, one can rewrite (3) in a more habitual form for the Reynolds equation [1-5]:

$$\frac{\partial \langle V_i \rangle}{\partial t} + \langle V_k \rangle \frac{\partial \langle V_i \rangle}{\partial x_k} = -\frac{1}{\rho}\frac{\partial \langle P \rangle}{\partial x_i} + \nu \frac{\partial^2 \langle V_i \rangle}{\partial x_k^2} - \frac{\partial \overline{(\delta V_i \delta V_k)}}{\partial x_k} \tag{4}$$

Of course, this equation should be analyzed together with the equation of continuity for mean velocity:

$$\frac{\partial \langle V_i \rangle}{\partial x_i} = 0 \tag{5}$$

Now it is necessary to solve the main problem: to express the last term in (4), containing the fluctuating velocity components $\overline{\delta V_i \delta V_k}$, through the mean velocity $\langle V_i \rangle$. To do this correctly, we will use the results obtained in modern theory of turbulence [4,6].

The most important part in any turbulent flow is played by the largest eddies (fluctuations): they have the largest dimensions and the largest velocity and pressure amplitudes [6, 7]. It means that only the large eddies influence significantly the mean characteristics of any turbulent flow (especially, the mean velocity and pressure distributions).

The small eddies participate in the turbulent flow with small velocity and pressure amplitudes. They may be regarded as a fine detailed structure superposed on the fundamental large turbulent eddies [6, 7].

The large eddies derive their kinetic energy from the average motion of the fluid (or gas) just because they have no other source of energy. So their energy space distribution will be non-homogeneous and similar to distribution in the mean flow.

However, a significant portion of the large eddies energy passes to the smaller eddies according to energy cascade and eventually dissipates in the smallest eddies.

In different areas of a turbulent flow there is a various number of smaller eddies. The greater is "the degree of turbulence" in the flow region under consideration, the more "generations" of smaller eddies will exist there. It is clear from the physical point of view that this "turbulence level" is determined by relative value of the mean velocity <V>.

So, we can assume that in a boundary layer (for small <V> values) due to the appreciable magnitudes of viscous forces the "degree of turbulence" is low, i.e. the influence of small eddies is insignificant and energy obtained by the large eddies from mean flow mainly remains within them. In the central area of a turbulent flow <V> values are large (i.e. "turbulence level" is high) and the numerous small eddies, existing there, "suck" energy from the large eddies.

As a result, the distribution of large eddies kinetic energy density $\overline{\delta V_i^2}$ in the flow will be a function growing from zero (at the wall of a channel) to some maximum (approximately at the boundary of viscous sub-layer) and smoothly decreasing to a constant level in central region of the flow.

Emphasize, that energy taken from the mean flow is equal [1-5] to the turbulent energy production, which, in its turn, is equal to the sum of three terms: viscous energy dissipation, diffusion of turbulent energy (turbulent transport) and convection (advection) of turbulent energy.

The Reynolds stress $\overline{\delta V_i \delta V_k}$ distribution in a turbulent flow should be similar to kinetic energy density $\overline{\delta V_i^2}$ distribution.

Now, assuming that we understand correctly the mechanism of turbulent motion and influence of large eddies on the mean flow parameters, we can express the Reynolds stress through the mean velocity. The most natural way to do it, is to expand the Reynolds stress into power series with respect to mean velocity and take a few first terms of this series:

$$\overline{\delta V_i \delta V_k} \approx D\langle V \rangle + A\langle V \rangle^2 + B\langle V \rangle^3 \quad , \tag{6}$$

where *A*, *B* and *D* are the constants.

It is obvious, that the Reynolds stress is equal to zero at the wall of channel (there are no velocity fluctuations at the wall) and in the center of flow (where two fluctuating velocity components $\delta V_i$ and $\delta V_k$ are not correlated [2, 4]). Between these points the Reynolds stress reaches some maximum (approximately at the boundary of viscous sub-layer [1-5]).

To satisfy these conditions we should write (6) as follows

$$\begin{cases} \overline{\delta V_i\, \delta V_k} = D\langle V\rangle + A\langle V\rangle^2 - B\langle V\rangle^3 \\ DV_{cl} + AV_{cl}^2 - BV_{cl}^3 = 0 \end{cases}, \tag{7}$$

where $V_{cl}$ is the mean velocity along centerline of a flow.

The formulae (7) could be simplified as

$$\overline{\delta V_i\, \delta V_k} = \left(BV_{cl}^2 - AV_{cl}\right)\langle V\rangle + A\langle V\rangle^2 - B\langle V\rangle^3, \tag{8}$$

and coefficients $A$ and $B$ will be determined below.

As a rule, the Reynolds stress (8) is positive and works against the mean velocity gradient removing energy from mean flow and passing it to the eddies. However, the Reynolds stress and correlation coefficient $\dfrac{\overline{\delta V_i\, \delta V_k}}{\sqrt{\overline{\delta V_i^2}}\sqrt{\overline{\delta V_k^2}}}$ sometimes can be negative [1, 4, 7], e.g. in the small areas of a turbulent flow close to channel wall. In such a region the mean flow gains kinetic energy from the eddies.

System of equations (4), (5), (8) describes, in principle, any turbulent flow. For some idealized cases it can be simplified and solved without particular difficulties.

## III. 1D STATIONARY TURBULENT FLOW IN CHANNEL AND PIPE

Begin our analysis with the simplest 1D case: turbulent stationary flow in a plane—parallel channel. We use the Cartesian coordinates $(x, y, z)$, with origin in the middle of channel. Assume that flow goes in $z$-direction, the height of channel in $x$-direction is $2a$ and the channel is infinite in $y$-direction. Since flow in such a channel is symmetric relative to $x=0$ plane, we will further analyze the velocity distribution only for $0 \le x \le a$.

For this stationary steady flow the first term in (4) is equal to zero, and the Reynolds stress, mean velocity $\langle V_z \rangle$ and mean pressure $\langle P \rangle$ do not depend on $z$-coordinate. Then, it follows from (5), that $\langle V_x \rangle = 0$.

Taking all this into account we obtain the following equations:

$$\frac{1}{\rho}\frac{\partial \langle P\rangle}{\partial x} + \frac{d\left(\overline{\delta V_x \delta V_x}\right)}{dx} = 0 \tag{9}$$

$$\frac{1}{\rho}\frac{\partial \langle P\rangle}{\partial z} - \nu\frac{d^2\langle V_z\rangle}{dx^2} + \frac{d\left(\overline{\delta V_x \delta V_z}\right)}{dx} = 0 \tag{10}$$

Since $\langle V_z \rangle$ does not depend on $z$, (10) can be satisfied only if

$$\frac{1}{\rho}\frac{\partial \langle P\rangle}{\partial z} = \nu\frac{d^2\langle V_z(x)\rangle}{dx^2} - \frac{d\left(\overline{\delta V_x \delta V_z}\right)}{dx} = -C, \tag{11}$$

where $C$ is the constant.

Using (8) and (11) we obtain the equation describing the mean velocity distribution for stationary turbulent flow in plane-parallel channel:

$$C + \nu\frac{d^2\langle V_z\rangle}{dx^2} = 2A\langle V_z\rangle\frac{d\langle V_z\rangle}{dx} - 3B\langle V_z\rangle^2\frac{d\langle V_z\rangle}{dx} + \left(BV_{cl}^2 - AV_{cl}\right)\frac{d\langle V_z\rangle}{dx} \tag{12}$$

Three terms in right-hand side in (12) describe a "turbulent contribution" in the mean velocity distribution. If we neglect the fluctuating velocities (the last term in (4)), it will reduce equation (12) to zero right-hand side, and this equation will describe the regular laminar flow in a channel.

Note that coefficients $A$ and $B$ in [12] determine the magnitude of convective acceleration terms. The right-hand side terms in (12) describe the "substantial" acceleration of fluid (or gas) particle that appears due to "specific" velocity and pressure distributions inside the turbulent flow.

Now we analyze the second simplest case: a turbulent stationary axi-symmetric flow in circular pipe with radius $R$. All the simplifications used above for channel are valid now. Applying them to equation (3) written in the cylindrical coordinates $(r, \varphi, z)$ for axi-symmetric case, we obtain the equation describing mean velocity distribution $\langle V_z(r) \rangle$ in 1D turbulent stationary flow

$$C + \nu \frac{d^2 \langle V_z \rangle}{d r^2} + \nu \frac{1}{r} \frac{d \langle V_z \rangle}{d r} = \\ 2 A \langle V_z \rangle \frac{d \langle V_z \rangle}{d r} - 3 B \langle V_z \rangle^2 \frac{d \langle V_z \rangle}{d r} + \left( B V_{cl}^2 - A V_{cl} \right) \frac{d \langle V_z \rangle}{d r} \quad (13)$$

Unlike (12), equation (13) contains one additional term in the left-hand side due to differentiation in cylindrical coordinates.

Equations (12) and (13) can be solved only numerically. To do this we should know coefficients $A$ and $B$, and centerline velocity $V_{cl}$. These quantities can be determined by specifying some characteristics of a turbulent flow and the boundary conditions: zero velocity at the wall, derivative at the wall and zero derivative at the center [5].

To specify a stationary fully developed turbulent 1D flow in channel or pipe we will use two parameters that can be measured without a difficulty: the bulk (average) velocity $V_b$ of flow and the pressure gradient (drop) $\frac{\partial \langle P \rangle}{\partial z}$ in axial direction. Besides this, one should, of course, know the kinematic viscosity $\nu$ and density $\rho$ of fluid (or gas) used, and channel half-height $a$ or pipe radius $R$.

Note, that in order to describe any laminar flow we need to specify only one measured parameter, because the quantities $V_b$, $V_{cl}$ and $\frac{\partial \langle P \rangle}{\partial z}$ for a laminar flow are not independent. For channel or pipe they are connected by the following formulas [1, 6, 7]:

$$V_b = \frac{2}{3} V_{cl} = \frac{a^2}{3 \nu \rho} \frac{\partial \langle P \rangle}{\partial z} \quad , \quad V_b = \frac{1}{2} V_{cl} = \frac{R^2}{8 \nu \rho} \frac{\partial \langle P \rangle}{\partial z} \quad , \quad (14)$$

Constant $C$ from (11), determining the pressure drop in axial direction, can be expressed through different parameters of a turbulent flow in channel and pipe, as follows [2-5, 8, 9]:

$$C = \frac{1}{\rho} \frac{\partial \langle P \rangle}{\partial z} = \left( \frac{\nu}{a} \frac{d \langle V_z \rangle}{dx} \right)_{x=a} = \frac{V_*^2}{a}, \quad C = \frac{1}{\rho} \frac{\partial \langle P \rangle}{\partial z} = \left( \frac{\nu}{R} \frac{d \langle V_z \rangle}{dr} \right)_{r=R} = \frac{V_*^2}{R} \quad (15)$$

where $V_*$ is the friction velocity.

The expressions for bulk velocities $V_b$ can be written in the following form for a channel and pipe, correspondingly:

$$V_b = \frac{1}{a}\int_0^a \langle V_z(x)\rangle\, dx = \frac{\text{Re}\,\boldsymbol{n}}{2a} \quad , \qquad V_b = \frac{2}{R^2}\int_0^R \langle V_z(r)\rangle\, r\, dr = \frac{\text{Re}\,\boldsymbol{n}}{2R} \quad , \qquad (16)$$

where the Reynolds number $\text{Re} = \dfrac{V_b D}{\boldsymbol{n}}$.

To find the desired mean velocity distributions we used the following computation method. Determining $V_b$ and $\dfrac{\partial \langle P\rangle}{\partial z}$ by the experimental data, specifying $\left.\dfrac{d\langle V_z\rangle}{dx}\right|_{x=0} = 0$ for a channel (or $\left.\dfrac{d\langle V_z\rangle}{dr}\right|_{r=0} = 0$ for a pipe), calculating coefficient $C$ by formulae (15), and using some concrete values for $A$ and $Vcl$ (e.g. $A = 0.01$ and $Vcl = 1.2Vb$), we solved equations (12) or (13) changing the value of coefficient $B$ until the obtained mean velocity profile would satisfy boundary conditions: $\langle V_z(x)\rangle|_{x=a} = 0$ for channel or $\langle V_z(r)\rangle|_{r=R} = 0$ for a pipe.

After that we calculated the bulk velocity by formula (16) and compared it with the experimental value. If this experimental result was greater (less) than the computed one, we increased (decreased) the magnitude of coefficient $A$ and repeated the calculations described above. This procedure has been done a few times until the obtained theoretical value $Vb$ coincides with the experimental data. Such a process is a converging one, and it usually takes only a few steps to achieve the match.

Then, using the calculated mean velocity profile ($\langle V_z(x)\rangle$ for a channel or $\langle V_z(r)\rangle$ for a pipe), we computed the derivative $\left.\dfrac{d\langle V_z\rangle}{dx}\right|_{x=a}$ at wall for a channel (or $\left.\dfrac{d\langle V_z\rangle}{dr}\right|_{r=R}$ for a pipe) and compared it with specified experimental value obtained with the help of measured pressure drop $\dfrac{\partial \langle P\rangle}{\partial z}$ and formula (15). If the experimental value was greater (less) than theoretical one, we decreased (increased) the magnitude of center line velocity $Vcl$ and repeated the whole process of calculations. Usually, the necessary agreement between theoretical and experimental data for velocity derivative at wall achieves after a few steps.

Thus, using equations (12) or (13) and varying in them parameters $A$, $B$, and $Vcl$, we determined their values, at which the desirable turbulent velocity profiles ($\langle V_z(x)\rangle$ for a channel or $\langle V_z(r)\rangle$ for a pipe) were obtained. These mean velocity distributions are solutions of equations (12) or (13), they give theoretical values for $V_b$ and $C = \dfrac{1}{\rho}\dfrac{\partial \langle P\rangle}{\partial z}$ coinciding with experimental ones, and satisfy the boundary conditions for a plane-parallel channel $\left.\dfrac{d\langle V_z\rangle}{dx}\right|_{x=0} = 0$, $\left.\dfrac{d\langle V_z\rangle}{dx}\right|_{x=a} = \dfrac{Ca}{\mathbf{n}}$, $\langle V_z\rangle|_{x=a} = 0$ or for a circular pipe $\left.\dfrac{d\langle V_z\rangle}{dr}\right|_{r=0} = 0$, $\left.\dfrac{d\langle V_z\rangle}{dr}\right|_{r=R} = \dfrac{CR}{\mathbf{n}}$, $\langle V_z\rangle|_{r=R} = 0$.

In general we compared almost forty different experimental velocity profiles taken from [3, 8-19], with our computations. Everywhere we obtained good agreement between our theoretical distributions and experimental data. Some examples for channels and pipes, for various fluids and gases with different Re are given in Figs. 1-10. As one can see, our calculations match perfectly well with the measurement results. It means that expression (8) for the Reynolds stress is really correct, i.e. we have found the right formula connecting this stress and mean flow velocity. In Table 1 we represented the main parameters of flows under consideration and calculated values of different coefficients.

Knowing mean velocity distributions $\langle V_z(x)\rangle$ for a channel or $\langle V_z(r)\rangle$ for a pipe and using formula (8), we are able to compute the Reynolds stress $S$:

$$S = \dfrac{\overline{dV_x dV_z}}{V_*^2} = \dfrac{(BV_{cl}^2 - AV_{cl})\langle V_z(x)\rangle + A\langle V_z(x)\rangle^2 - B\langle V_z(x)\rangle^3}{V_*^2}$$
$$S = \dfrac{\overline{dV_r dV_z}}{V_*^2} = \dfrac{(BV_{cl}^2 - AV_{cl})\langle V_z(r)\rangle + A\langle V_z(r)\rangle^2 - B\langle V_z(r)\rangle^3}{V_*^2}$$
(17)

These quantities were calculated for seven cases described in [8, 9, 11, 19] and compared with corresponding experimental data. Typical results given in Figs. 11-14 show a good match between theoretical and experimental distributions, which once again confirms the correctness of our approach and accuracy of calculations.

## IV. CONCLUSIONS

Mean velocity distributions for turbulent flows can be approximately described by equations (4), (5), (8). This approach is based on relation (8) between the Reynolds stress and mean velocity components in a turbulent flow.

Computation results for two simplest 1D stationary fully developed turbulent flows in a circular pipe and a plane-parallel channel, based on equations (12), (13), match perfectly well with experimental data for the mean velocity and Reynolds stress distributions.

The agreement between theoretical and experimental results confirms the correctness of our approach and computation accuracy.

## ACKNOWLEDGEMENTS

The author is the most grateful to Dr. R. S. C. Cobbold and Dr. P. A. J. Bascom at the Institute of Biomedical Engineering, University of Toronto for their significant contribution, help, suggestions, discussion, and encouragement.

# TABLE 1

| Refe-rence | Re *10³ | Medium and its viscosity, (mm²/s) | Coeffi-cient A*10³ | Coeffi-cient B*10⁶ (s/mm) | Coeffi-cient C (mm/s²) | Condi-tions | Pipe diameter D or channel height 2a in mm | Vbulk (m/s) |
|---|---|---|---|---|---|---|---|---|
| [10] | 8 | Air, 15 | 4.0 | 2.77 | 3310 | Normal | D=33 | 2.545 |
|  | 50 | Air, 15 | 2.0 | 0.4079 | 105570 | Normal | D=33 | 17.46 |
| [8] | 40.2 | Air, 15 | 15 | 7.827 | 440 | Normal | D=247 | 2.44 |
|  | 428 | Air, 15 | 25 | 0.80923 | 26940 | Normal | D=247 | 25.51 |
| [9] | 21.4 | Air, 15 | 15 | 4.23 | 200 | Normal | 2a=127 | 2.53 |
|  | 55.4 | Air, 15 | 18.7 | 2.007 | 1191 | Normal | 2a=127 | 6.55 |
|  | 113 | Air, 15 | 17 | 0.9329 | 4547 | Normal | 2a=127 | 13.35 |
| [18] | 4 | Water, 1.35 | 60 | 71.43 | 594 | T=90 C | D=10 | 0.545 |
|  | 23 | Water, 1.35 | 25 | 6.57 | 12410 | T=90 C | D=10 | 3.15 |
|  | 110 | Water, 1.14 | 17 | 3.678 | 4533 | T=150 C | D=30 | 4.0 |
|  | 1100 | Water, 1.12 | 20 | 1.2081 | 8525 | T=170 C | D=100 | 12.45 |
|  | 2000 | Water, 1.1 | 25 | 0.8077 | 22900 | T=190 C | D=100 | 21.5 |
|  | 3200 | Water, 0.75 | 35 | 0.934 | 29520 | T=380 C | D=100 | 24.3 |
| [16] | 7000 | Natural gas, 0.2425 | 20 | 0.7877 | 11020 | Pressure 5200 kPa | D=102.26 | 16.6 |
| [17] | 4.65 | Oil, 6 | 30 | 204.5 | 1.3 | Normal | 2a=220 | 0.127 |
| [12] | 7.00 | Oil, 6 | 40 | 166.7 | 2.65 | Normal | 2a=220 | 0.191 |
| [14] | 11.8 | Water, 1 | 15 | 103.9 | 3 | Normal | 2a=80 | 0.149 |
|  | 29.8 | Water 1 | 23 | 52.66 | 22.3 | Normal | 2a=80 | 0.397 |
|  | 42.2 | Water, 1 | 30 | 42.76 | 31.3 | Normal | 2a=80 | 0.53 |
| [11] | 5.74 | Water, 1 | 55 | 323.4 | 1.89 | Normal | 2a=48.8 | 0.102 |
| [19] | 36.7 | Water, 1 | 20 | 22.19 | 46.59 | Normal | 2a=48.8 | 0.663 |
| [15] | 26 | Sugar solution, 37.3 | 15 | 1.361 | 66535 | Normal | D=25.4 | 3.812 |
|  | 13.4 | Sugar solution, 17.4 | 13 | 0.96 | 25870 | Normal | D=25.4 | 9.363 |
|  | 91.7 | Water, 1 | 22 | 5.526 | 1789 | Normal | D=25.4 | 2.79 |

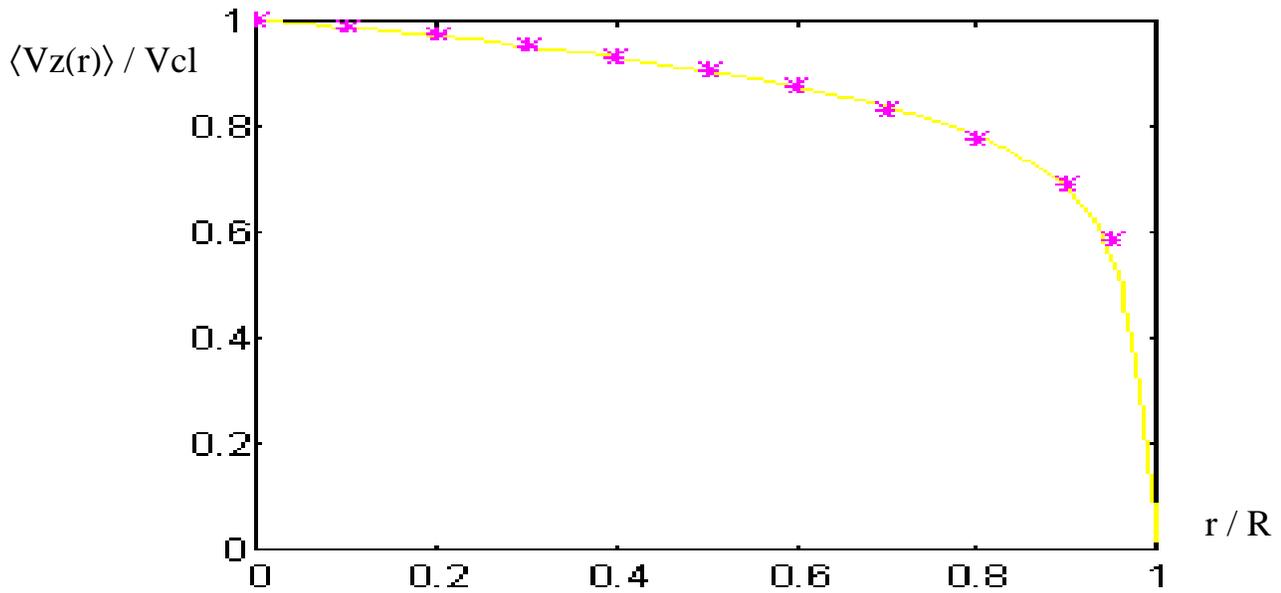

Figure 1. Mean velocity profile (in relative units) vs. pipe radius (in relative units) for water flow (Re=4000, $\nu$=1.35 mm$^2$/s, $V_b$=0.545 m/s, pipe diameter 10 mm). Stars are experimental data from [18], solid line is our theoretical distribution.

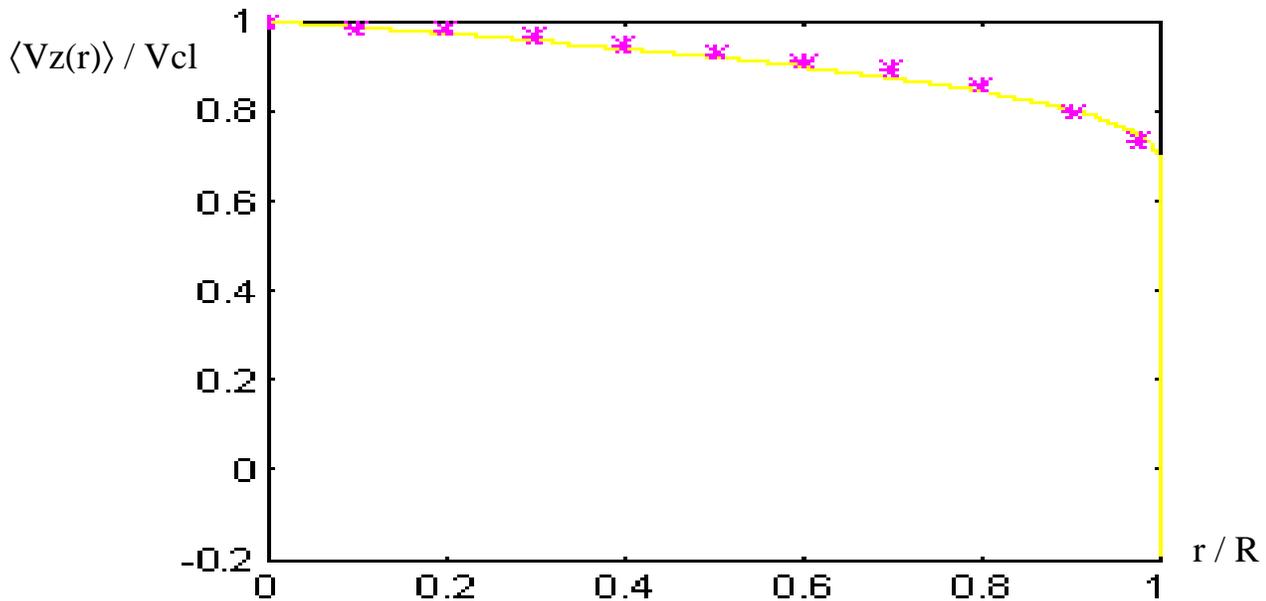

Figure 2. Mean velocity profile (in relative units) vs. pipe radius (in relative units) for water flow (Re=3200000, $\nu$=0.75 mm$^2$/s, $V_b$=24.3 m/s, pipe diameter 100 mm). Stars are experimental data from [18], solid line is our theoretical distribution.

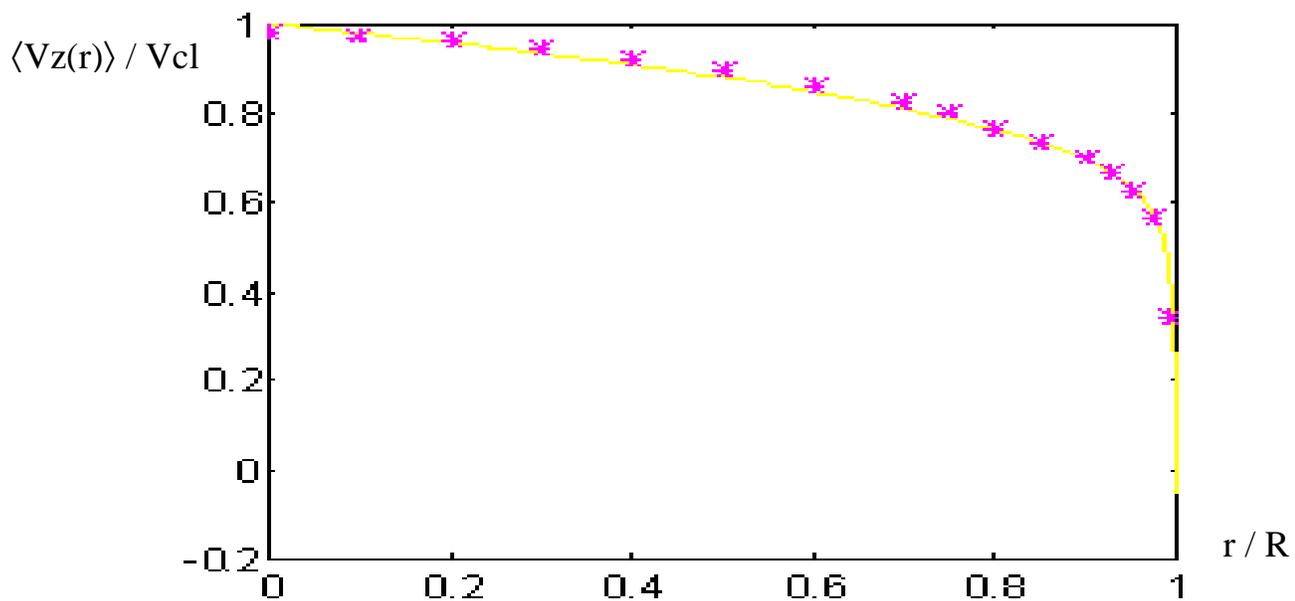

Figure 3. Mean velocity profile (in relative units) vs. pipe radius (in relative units) for Air flow (Re=40200, $\nu$=15 mm$^2$/s, $V_b$=2.44 m/s, pipe diameter 247 mm). Stars are experimental data from [8], solid line is our theoretical distribution.

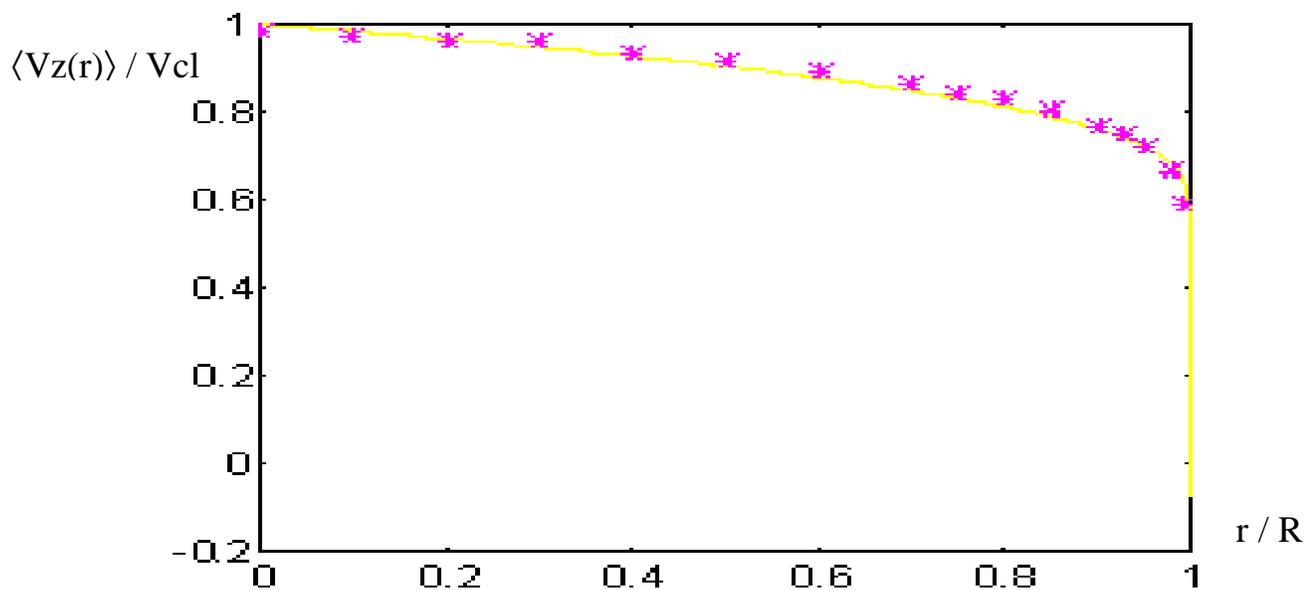

Figure 4. Mean velocity profile (in relative units) vs. pipe radius (in relative units) for Air flow (Re=428000, $\nu$=15 mm$^2$/s, $V_b$=25.51 m/s, pipe diameter 247 mm). Stars are experimental data from [8], solid line is our theoretical distribution.

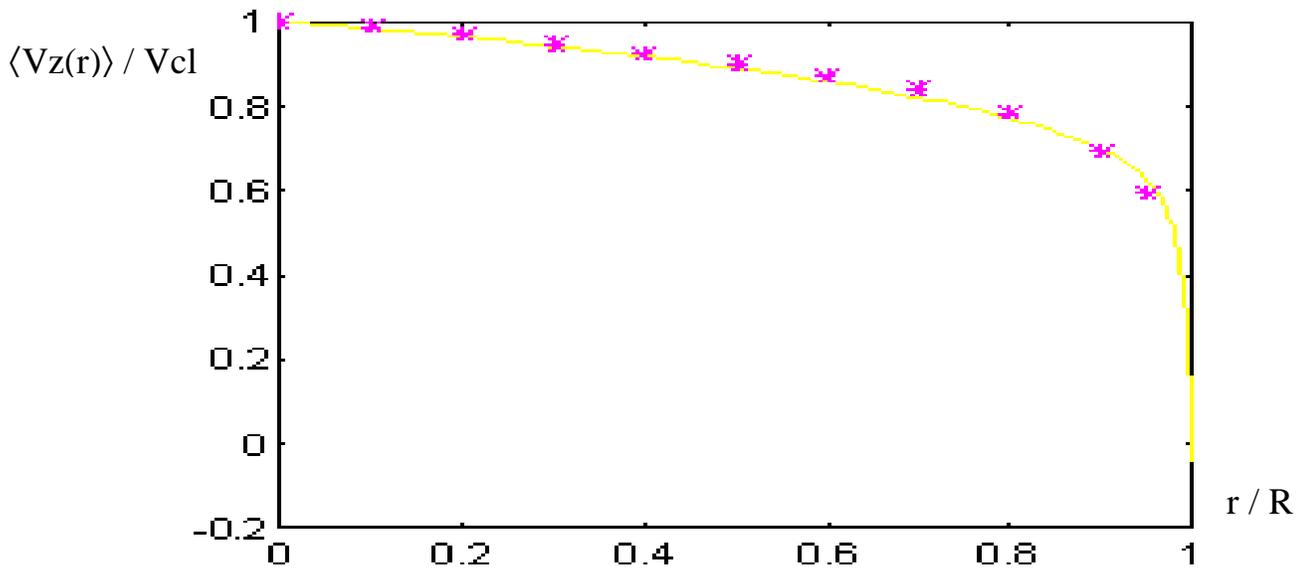

Figure 5. Mean velocity profile (in relative units) vs. pipe radius (in relative units) for flow of sugar solution (Re=26000, $\nu$=37.3 mm$^2$/s, $V_b$=3.812 m/s, pipe diameter 25.4 mm). Stars are experimental data from [15], solid line is our theoretical distribution.

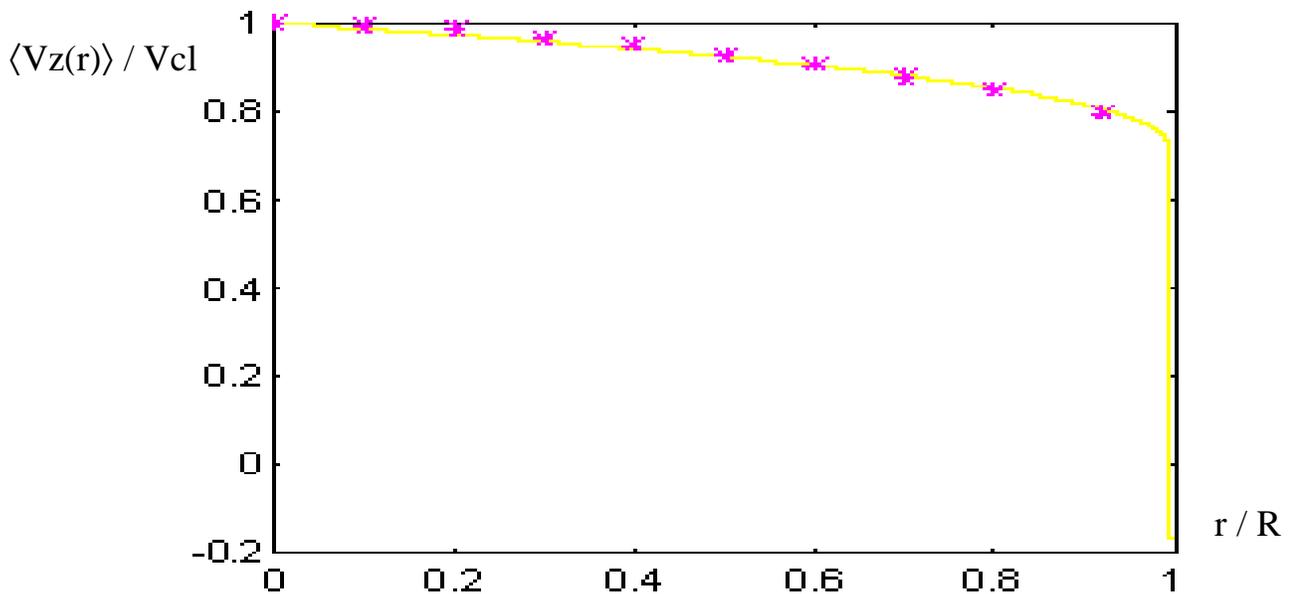

Figure 6. Mean velocity profile (in relative units) vs. pipe radius (in relative units) for flow of natural gas (Re=7000000, $\nu$=0.2425 mm$^2$/s, $V_b$=16.6 m/s, pipe diameter 102.26 mm). Stars are experimental data from [16], solid line is our theoretical distribution.

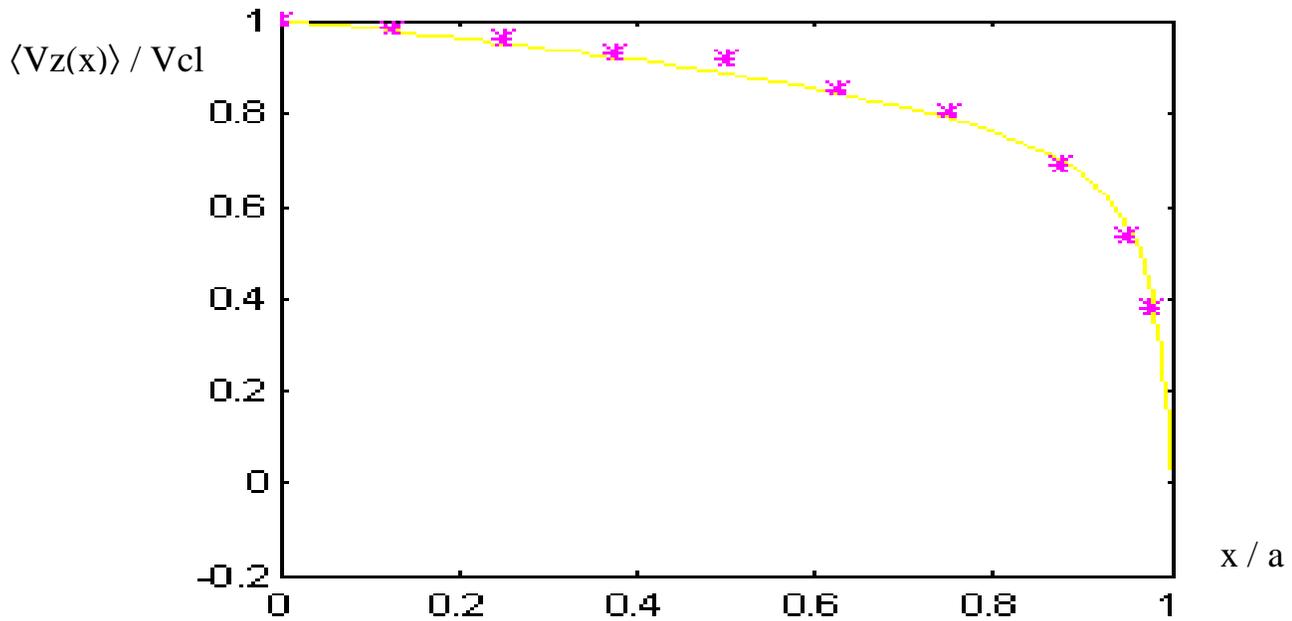

Figure 7. Mean velocity profile (in relative units) vs. channel height (in relative units) for water flow (Re=5740, $\nu=1$ mm$^2$/s, $V_b$=0.102 m/s, channel height 48.8 mm). Stars are experimental data from [11], solid line is our theoretical distribution.

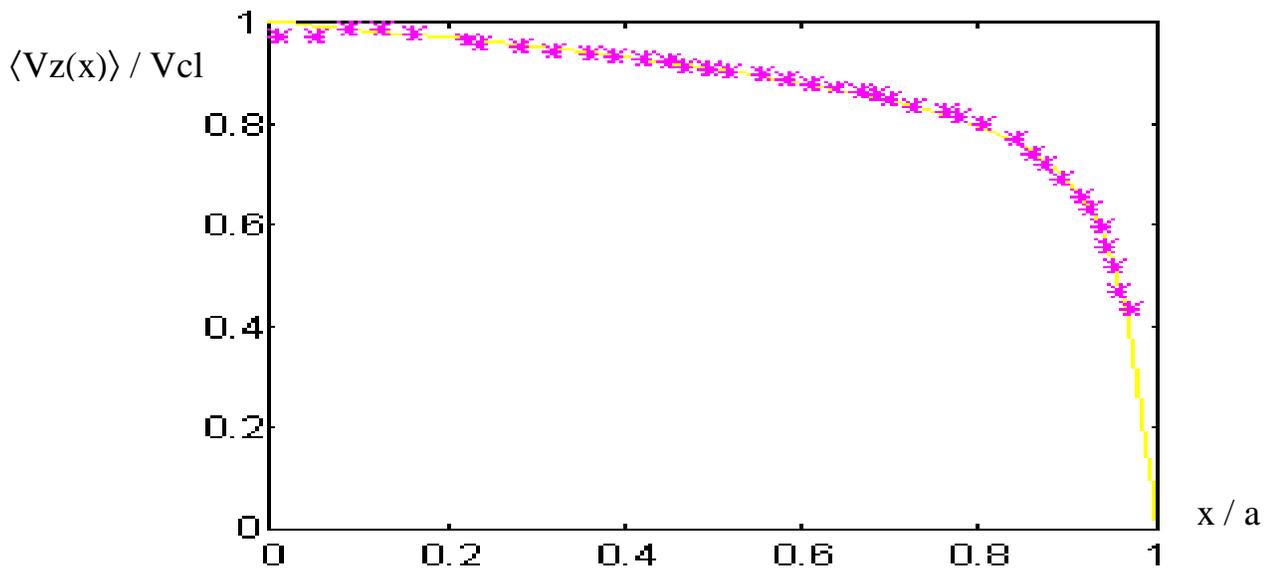

Figure 8. Mean velocity profile (in relative units) vs. channel height (in relative units) for water flow (Re=36700, $\nu=1$ mm$^2$/s, $V_b$=0.663 m/s, channel height 48.8 mm). Stars are experimental data from [19], solid line is our theoretical distribution.

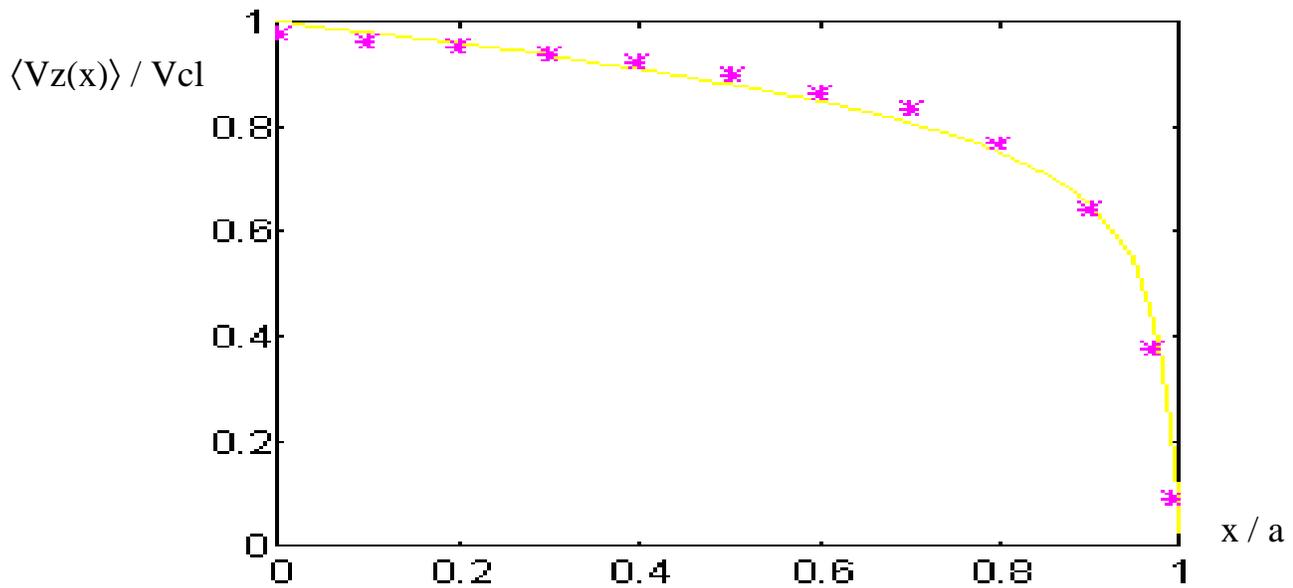

Figure 9. Mean velocity profile (in relative units) vs. channel height (in relative units) for oil flow (Re=4650, $\nu$=6 mm$^2$/s, $V_b$=0.127 m/s, channel height 220 mm). Stars are experimental data from [17], solid line is our theoretical distribution.

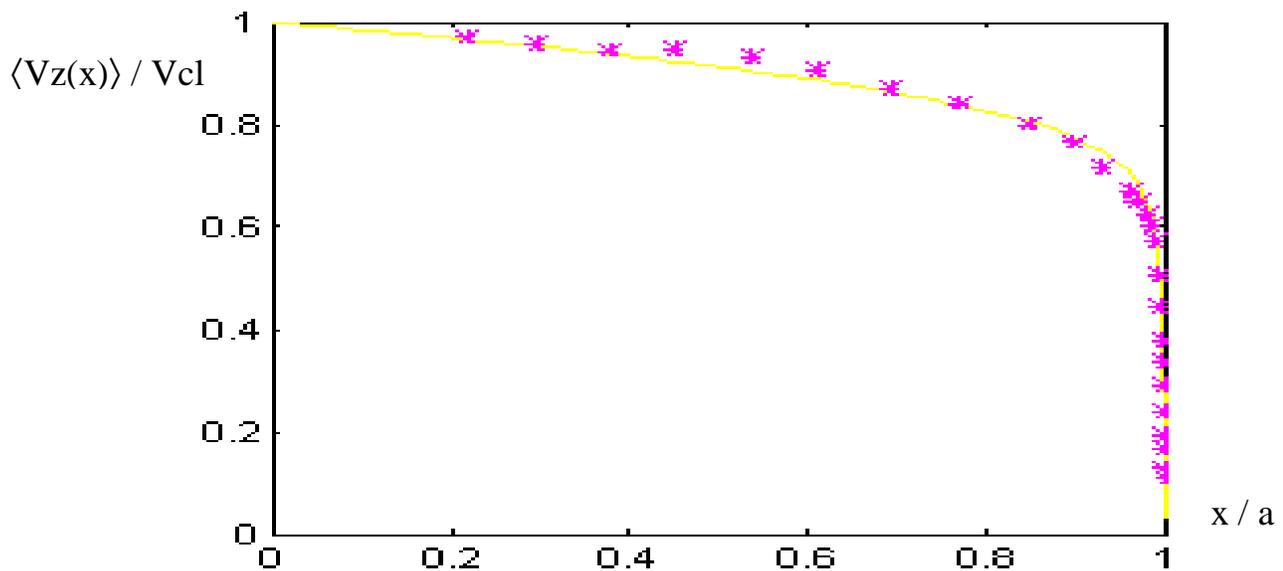

Figure 10. Mean velocity profile (in relative units) vs. channel height (in relative units) for air flow (Re=113000, $\nu$=15 mm$^2$/s, $V_b$=13.35 m/s, channel height 127 mm). Stars are experimental data from [9], solid line is our theoretical distribution.

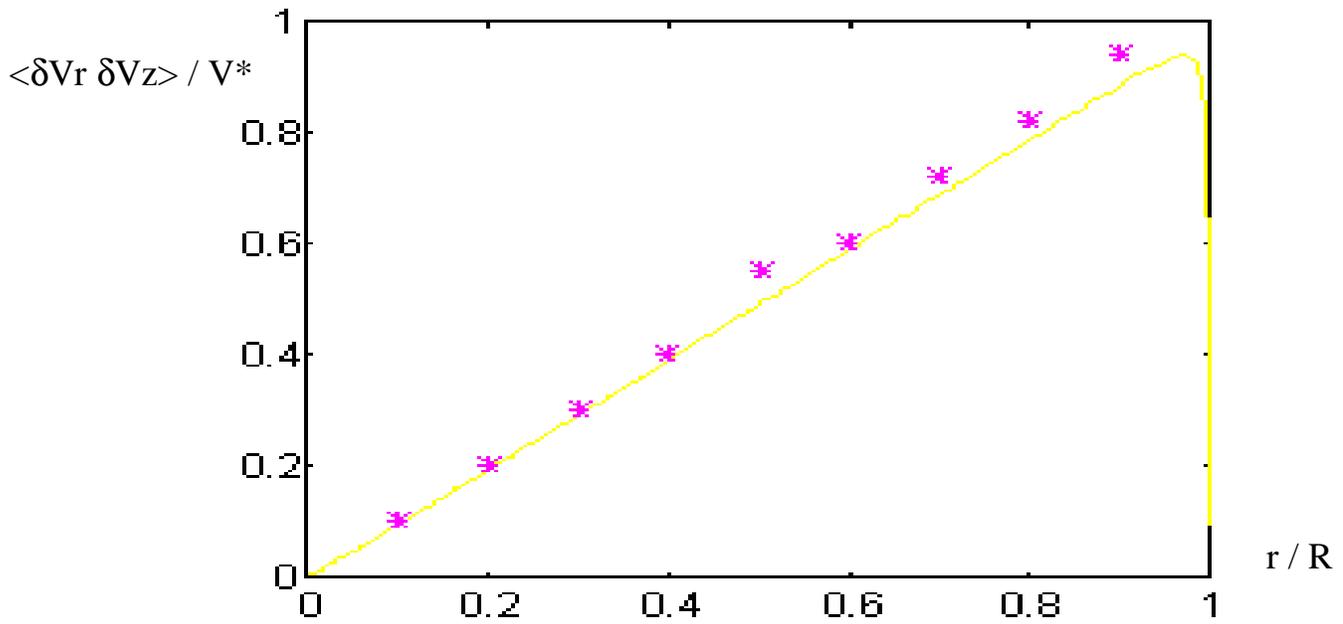

Figure 11. Reynolds stress distribution (in relative units) vs. pipe radius (in relative units) for air flow (Re=40200, $\nu$=15 mm$^2$/s, $V_b$=2.44 m/s, pipe diameter 247 mm). Stars are experimental data from [8], solid line is our theoretical distribution.

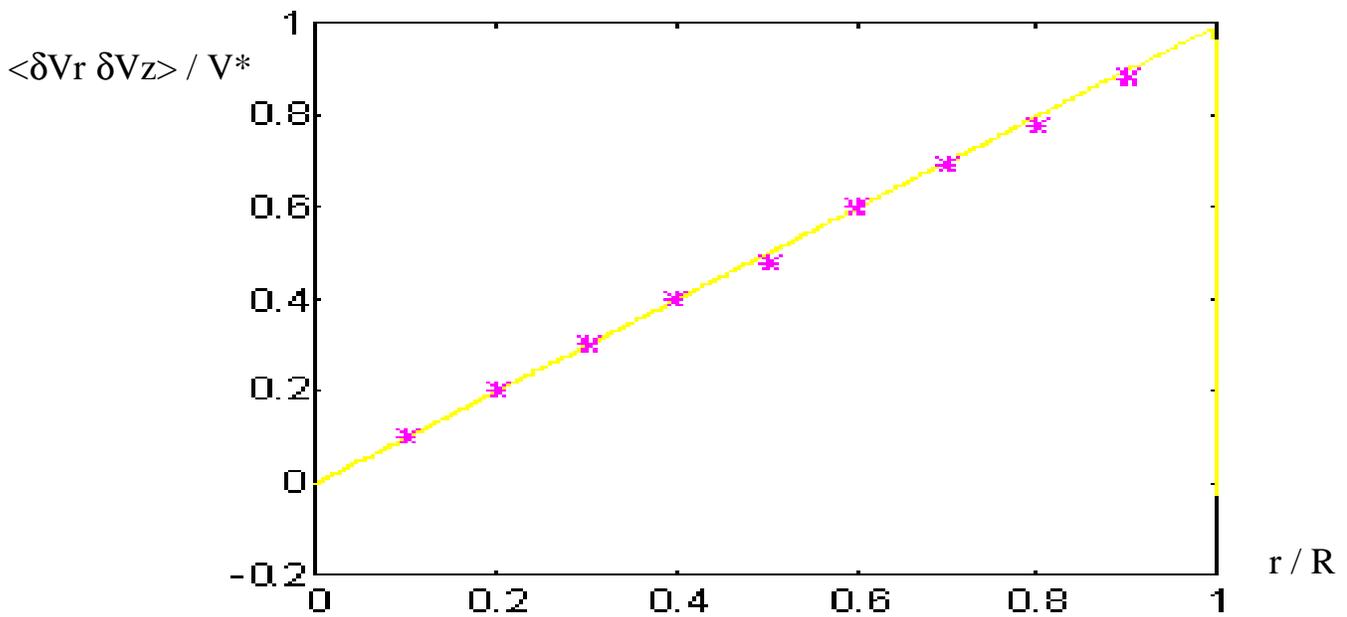

Figure 12. Reynolds stress distribution (in relative units) vs. pipe radius (in relative units) for air flow (Re=428000, $\nu$=15 mm$^2$/s, $V_b$=25.51 m/s, pipe diameter 247 mm). Stars are experimental data from [8], solid line is our theoretical distribution.

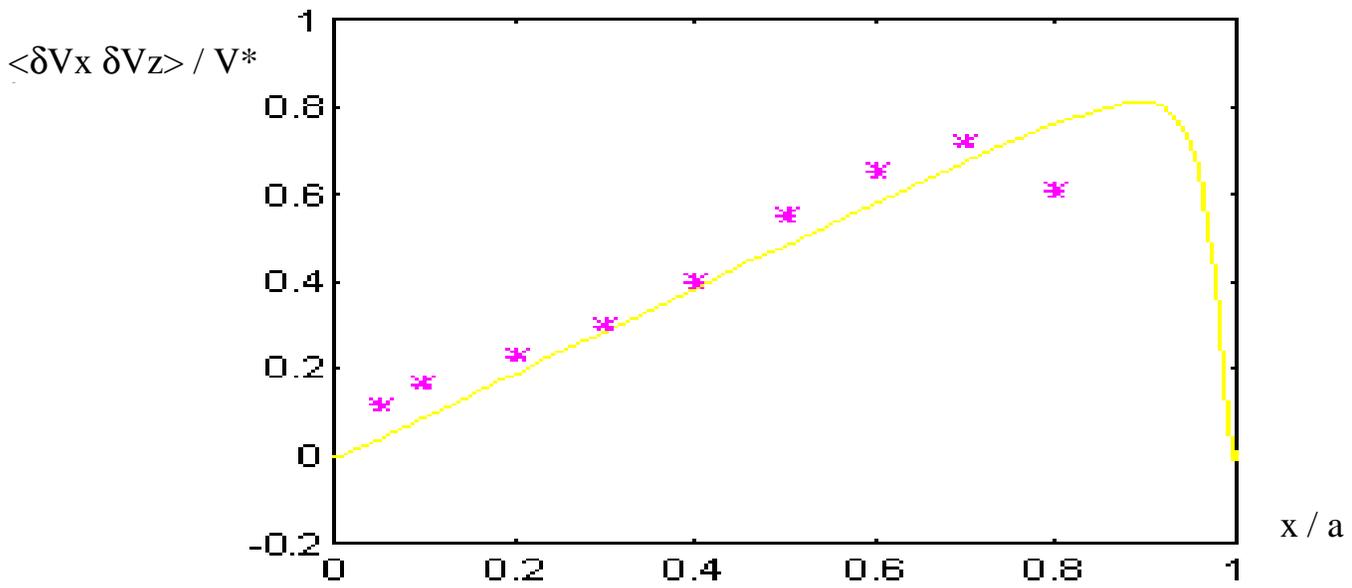

Figure 13. Reynolds stress distribution (in relative units) vs. channel height (in relative units) for air flow (Re=21400, $\nu$=15 mm$^2$/s, $V_b$=2.53 m/s, channel height 127 mm). Stars are experimental data from [9], solid line is our theoretical distribution.

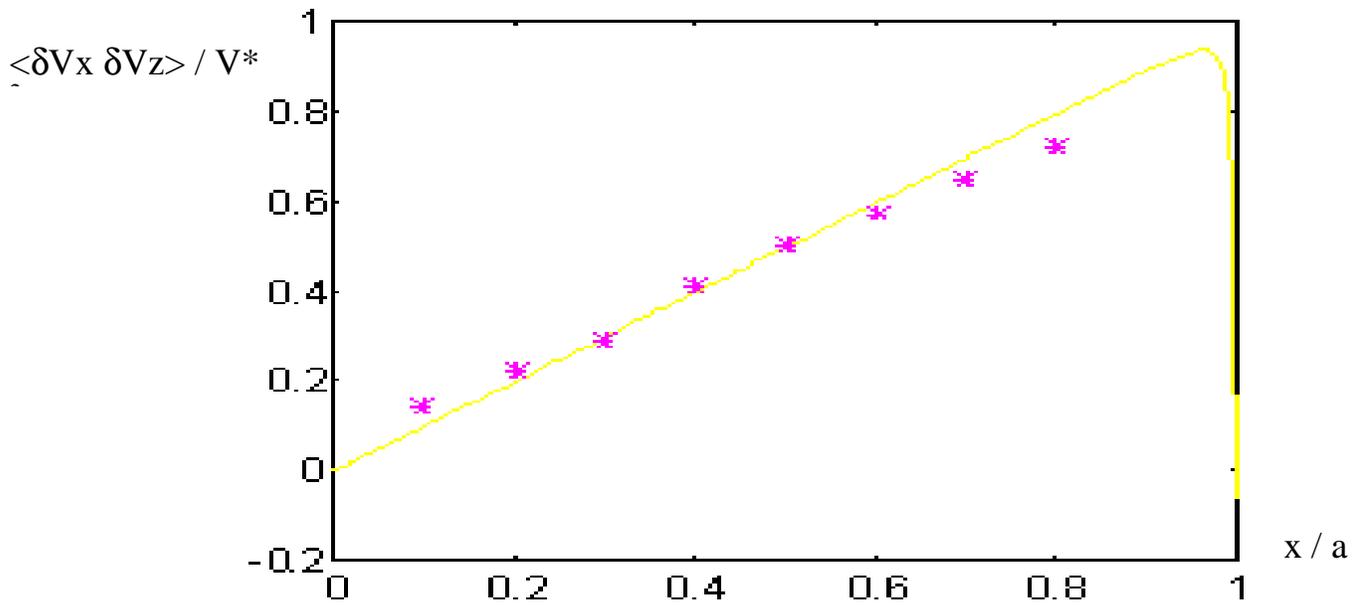

Figure 14. Reynolds stress distribution (in relative units) vs. channel height (in relative units) for air flow (Re=113000, $\nu$=15 mm$^2$/s, $V_b$=13.35 m/s, channel height 127 mm). Stars are experimental data from [9], solid line is our theoretical distribution.